\documentclass{elsart}
\usepackage{elsfix}
\usepackage{graphicx}            
\newcommand{ \qnlsm } {QNL$\sigma$M}
\newcommand{ \lco } {La$_{2}$CuO$_{4}$}
\newcommand{ \lbco } {(La,Ba)$_{2}$CuO$_{4}$}
\newcommand{ \lsnco } {(La,Sr,Nd)$_{2}$CuO$_{4}$}
\newcommand{ \lczmo } {La$_{2}$Cu$_{\rm 1-z}$(Zn,Mg)$_{\rm z}$O$_{4}$}

\newcommand{ \neel } {N\'{e}el}

\begin{document}

\bibliographystyle{elsart-num}

\begin{frontmatter}
\title{Neutron Scattering, Magnetometry, and Quantum Monte Carlo Study of
the Randomly-Diluted Spin-1/2 Square-Lattice Heisenberg Antiferromagnet}
\author[STANF1]{O.P. Vajk},
\author[STANF2,STANF3]{\underline{M. Greven}},
\author[STANF2]{P.K. Mang},
\and \author[NIST]{J.W. Lynn}

\address[STANF1]{Department of Physics, Stanford University, Stanford, CA 94305, USA}
\address[STANF2]{Department of Applied Physics, Stanford University, Stanford, CA 94305, USA}
\address[STANF3]{Stanford Synchrotron Radiation Laboratory,
                 Stanford University, Stanford, CA 94309, USA}
\address[NIST]{NIST Center for Neutron Research, National Institute of Standards and Technology,
                 Gaithersburg, MD 20899, USA}

\begin{abstract}

We have successfully grown sizable single crystals of \lczmo~with
up to nearly half of the magnetic Cu sites replaced by
non-magnetic Zn and Mg. Neutron scattering, SQUID magnetometry,
and complementary quantum Monte Carlo (QMC) simulations
demonstrate that this material is an excellent model system for
the study of site percolation of the square-lattice Heisenberg
antiferromagnet (SLHAF) in the quantum-spin limit $S=1/2$.
Carefully oxygen-reduced samples exhibit N\'eel order up to the
percolation threshold for site dilution, $z_p \approx 40.7 \%$. For
$z>10\%$, the material exhibits a low-temperature tetragonal (LTT)
structural phase, with a transition temperature that increases
linearly with doping. Above $z \approx 25\%$, N\'eel order occurs
in the LTT phase. Up to at least $z = 35\%$, the N\'eel
temperature $T_N (z)$ of the experimental system corresponds to
the temperature at which QMC indicates that the spin correlations
for the nearest-neighbor $S=1/2$ SLHAF have grown to approximately
100 lattice constants. Neutron scattering measurements of the
static structure factor in the paramagnetic regime allow the
determination of the two-dimensional spin correlations, which are
found to be in excellent quantitative agreement with QMC over a
wide common temperature and doping
range. Neutron scattering and QMC results
for the temperature dependence of the static structure factor
amplitude $S(\pi,\pi)$ are in good
agreement as well. As the concentration of non-magnetic sites is
increased, the magnetic correlation length $\xi (T,z)$ crosses over
from an exponential dependence on $\rho_s/T$ to power-law behavior
in the temperature regime studied. Fits to a heuristic crossover
form for $\xi (T,z)$ allow an estimate of the spin stiffness,
$\rho_s = \rho_s (z)$, which approaches zero at $z = z_p$. The
combined experimental and numerical data presented here provide
valuable quantitative information for tests of theories of the
randomly-diluted $S=1/2$ SLHAF.
\end{abstract}
\begin{keyword}
Magnetic dilution, two-dimensional, Heisenberg antiferromagnet,
quantum phase transition, neutron scattering, quantum Monte Carlo,
percolation
\end{keyword}
\end{frontmatter}

\underline{Corresponding author:} \\
M. Greven, T.H. Geballe Laboratory for Advanced Materials,
Stanford University, Stanford, CA 94305-4045.  Tel. 650-725-8978,
FAX: 650-724-3681, email: greven@stanford.edu


The discovery of high-temperature superconductivity in
charge-carrier doped La$_2$CuO$_4$ has stimulated an enormous
interest in the properties of low-dimensional quantum magnets. The
main structural building block of La$_2$CuO$_4$ and of related
Mott insulators is a square lattice of Cu$^{2+}$ spin-1/2 magnetic
ions that experience a strong nearest-neighbor (NN)
antiferromagnetic superexchange ($J
\approx 1550$ K) mediated by intervening oxygens.
Apart from small correction terms
(such as anisotropies and a three-dimensional coupling), which
lead to N\'eel order at a non-zero temperature,
the magnetic degrees of freedom of these CuO$_2$
sheets are well-described by the square-lattice Heisenberg
Hamiltonian
\begin{eqnarray}
\label{pure_hamiltonian}
\mathcal{H} = & J & \;
\sum_{\langle i, j \rangle} {\bf S}_i \cdot{\bf S}_{j},
\end{eqnarray}
where the sum is over NN sites, $J$ is the antiferromagnetic Cu-O-Cu superexchange,
and ${\bf S}_i$ is the $S=1/2$ operator at the site $i$.

Extensive experimental
\cite{lyons88,endoh88,aeppli89,tokura90,keimer92,greven95,birgeneau99,kim01,kim01L,coldea01},
quantum Monte Carlo
\cite{makivic91,kim98,beard98}, and theoretical
\cite{chakravarty89,hasenfratz91,tyc89,singh89,igarashi92,chubukov94,hamer94,elstner95,singh95,cuccoli97,hasenfratz00}
efforts have led to a good understanding of the quantum many-body physics
described by Eq. \ref{pure_hamiltonian}.
The discrete Hamiltonian Eq. (1) has been mapped to a continuum quantum
non-linear $\sigma$ model (QNL$\sigma$M), and it has been established
that the NN square-lattice
Heisenberg antiferromagnet (SLHAF) exhibits a broken-symmetry
ground state (and renormalized classical behavior) rather than a
quantum-disordered ground state even in the extreme quantum-spin limit
of spin-1/2 \cite{chakravarty89}.  Numerical and theoretical studies
indicate that the quantum critical point separating these two
ground states may be traversed, for example, upon the introduction
of frustrating next-NN interactions \cite{chakravarty89,siurakshina01} or by
coupling two SLHAFs to form a bilayer \cite{shevchenko00}.
Although the full spin Hamiltonian of \lco~is believed to contain
a frustrating next-NN exchange of about 0.05-0.10$J$
\cite{birgeneau99,kim01,annett89}, this is well below the value of
$\approx 0.24J$ needed to disorder the system \cite{siurakshina01}.

Quantum phase transitions in the presence of disorder are
the subject of considerable current interest. Magnetic systems are
often of particular value in this context, as the combined effects
of quantum fluctuations and quenched disorder can be studied with
various analytical and numerical methods. Furthermore, model
magnets, if they can be found in nature, can serve as an important
testing ground of theoretical predictions. A significant amount of
theoretical and numerical effort has been devoted to the relatively simple one-dimensional
\cite{young96,sachdev97} and two-dimensional \cite{senthil96,heuer92} Ising
model in a transverse field. For the $O(3)$ symmetric $S=1/2$
SLHAF, the effects of a single impurity are well understood
\cite{sandvik97,sachdev99,vojta00}, but there are few theoretical
results for finite impurity concentrations \cite{chernyshev02}.
There have been suggestions that random site
\cite{senthil96,yasuda99,chen00} and bond \cite{sandvik95}
dilution of the NN $S=1/2$ SLHAF may
lead to a non-trivial quantum phase transition.
Extensive experimental results exist
for $S=5/2$ up to very high impurity concentrations
\cite{birgeneau84}, but results for the extreme
quantum-spin limit of S=1/2 have been limited to lower concentrations
\cite{keimer92,chakraborty89,cheong91,ting92,lichti91,cao94,corti95,uchinokura95,hucker99,clarke95}.
In studies of site-diluted $S=5/2$ Heisenberg and Ising antiferromagnets
\cite{birgeneau84,breed70},
long-range order was found to disappear only above the percolation threshold
$z_p \approx 40.725\%$ \cite{Stauffer,newman00}.
Earlier results for the \neel~temperature of randomly-diluted
\lco~extrapolate to zero temperature at concentrations
well below the percolation threshold
\cite{keimer92,chakraborty89,cheong91,ting92,lichti91,cao94,corti95,uchinokura95,hucker99}.
These results, along with theoretical predictions
\cite{senthil96,yasuda99,chen00} and numerical studies \cite{miyashita92},
suggested the possible existence of a
new quantum critical point at $z_{S=1/2} < z_p$.
However, recent Monte Carlo simulations
\cite{kato00,todo01,yasuda01,vajk02,sandvik02,sandvik02b,vajk02b}
indicate that the site diluted NN $S=1/2$ SLHAF
remains ordered up to the percolation threshold.
While the zero-temperature transition appears to be a classical percolation
transition, the intermediate-temperature properties of the spin-1/2 system
at $z=z_p$ seem to be controlled by the effective
proximity to a new multicritical
point \cite{sandvik02b,vajk02b}

Very recently, we have been able to show that \lczmo~is a model
material for the study of the effects of nonmagnetic impurities in
the $S=1/2$ SLHAF up to and beyond the site percolation threshold
\cite{vajk02}.
Figure 1 is a schematic of the randomly diluted CuO$_2$ sheets in \lczmo,
showing the geometric transition from an infinite connected cluster
to finite disconnected clusters as the percolation threshold is traversed.
We have also performed quantum Monte Carlo (QMC) simulations of
the randomly diluted NN $S=1/2$ SLHAF,
\begin{eqnarray}
\label{hamiltonian}
\mathcal{H} = & J & \;
\sum_{\langle i, j \rangle} p_i p_j {\bf S}_i \cdot{\bf S}_{j},
\end{eqnarray}
where $p_i = 1$ ($p_i = 0$) on magnetic (nonmagnetic) sites.
The experimental results for the static structure factor in the
2D correlated parramagnetic region agree remarkably well with our
numerics \cite{vajk02}, indicating that \lczmo~is indeed a very
good model system.  The present paper outlines some
aspects of these results in more detail.

\subsection{Sample Growth and Characterization}

A lack of stoichiometric samples at high non-magnetic
concentrations had limited earlier experimental work on
randomly diluted \lco. In previous studies of polycrystalline
samples the concentration of non-magnetic ions was as high as 25\%
\cite{hucker99}, significantly below the percolation threshold.
Single crystal results had been limited even further, to $z
\approx 15$\% and below
\cite{keimer92,cao94,uchinokura95}.
Furthermore, the excess oxygen typically found in as-grown samples
introduces holes into the copper-oxygen sheets which frustrate
the antiferromagnetism and quickly destroy magnetic order
\cite{aharony88}. Differing values for $T_N (z)$ indicate that
this problem had not been fully resolved
\cite{keimer92,chakraborty89,cheong91,ting92,lichti91,cao94,corti95,uchinokura95,hucker99}.

We have successfully grown single crystals of \lczmo~by the
optical traveling-solvent floating-zone method.  Zn$^{2+}$ and
Mg$^{2+}$ are non-magnetic and have larger and smaller ionic radii, respectively,
than Cu$^{2+}$. By jointly substituting Zn and Mg for Cu, we succeeded
in growing samples up to and beyond the percolation threshold
\cite{vajk02}. The Zn content of our samples is approximately 10\%,
while the Mg content varies.  Typical single-grain sections
are about 4 mm in diameter and 40 mm long. The composition of the
samples was measured with electron probe microanalysis on sections
cut from the ends of the crystals.  Some crystals have small Cu/Zn/Mg
concentration gradients along the length ($\Delta z$ = 1-2\%),
but all crystals are very uniform radially. The mosaic
widths are very good, 15' full width at half maximum (FWHM) or
less, as measured by neutron diffraction. Samples were carefully
annealed for 24 hours at 900 to 950 $^{\circ}$C in an Ar
atmosphere to remove excess oxygen.  Using SQUID magnetometry, we
checked that subsequent anneals do not further raise $T_N$,
confirming that the first anneal was successful.

The neutron scattering measurements presented here were performed at the
National Institure of Standards and Technology (NIST) Center for
Neutron Research  (NCNR), using the BT2 and BT9 thermal
instruments and the SPINS cold neutron instrument.

\subsection{Phase Diagram}

Near the \neel~transition, the uniform susceptibility of
\lczmo~becomes coupled to the staggered susceptibility through the
antisymmetric \linebreak Dzyaloshinskii-Moriya term (not included
in Eqs. \ref{pure_hamiltonian} and \ref{hamiltonian}), and the uniform susceptibility
perpendicular to the CuO$_2$ sheets increases to a cusp at $T_N$
\cite{uchinokura95,thio88}. In order to determine the
\neel~temperature, small pieces, a few mm on each side,
were cut from the larger crystals for magnetometry. Figure 2a
shows the uniform susceptibility for several samples measured with
a SQUID magnetometer. Polycrystalline samples, with
concentrations assumed to be equal to their nominal values,
yielded similar results.  At higher dilution levels,
a large Curie-like component emerges
\cite{uchinokura95}, and the cusp at $T_N$ becomes more
difficult to distinguish.

Neutron diffraction measurements allow us to follow $T_N (z)$ at
higher concentrations. The intensity at a magnetic Bragg
reflection is proportional to the square of the staggered magnetic
moment, and the doping dependence of the extracted ordered moment
\cite{vajk02} agrees rather well with recent theoretical \cite{chernyshev02}
and numerical \cite{sandvik02} results for the NN SLHAF.
Figure 2b shows the temperature dependence of the magnetic
scattering at the (1,0,0) magnetic reflection for several samples
(we use orthorhombic ({\it Bmab}) notation throughout).
The neutron diffraction results for $T_N$
agree with SQUID measurements at low and intermediate
concentrations and indicate that \neel~order persists up to at
least $z=39\%$.

We typically observe an additional temperature-independent
non-magnetic signal which was subtracted from the data in Fig. 2b.
This remnant signal is a combination of double scattering from the
(1,0,2) and (0,0,2) nuclear Bragg reflections and a small $\lambda/2$
contamination due to nuclear scattering of higher-energy neutrons.
The double-scattering component is strongly energy-dependent,
and decreases at lower neutron energies.  Neutrons with half the
wavelength (and four times the energy) of the primary beam will
also be diffracted by the monochromator.
These neutrons are filtered out with pyrolytic graphite on thermal
instruments or berylium on SPINS, but small numbers may still
pass through if the filter is not sufficiently thick,
giving (2,0,0) structural Bragg scattering at the
same angular position as the primary beam at the (1,0,0) magnetic
peak.

The correction terms to Eqs. \ref{pure_hamiltonian}
and \ref{hamiltonian} are believed to
include a small frustrating next-NN exchange
\cite{birgeneau99,kim01,annett89}, which might in principle become more
important at higher concentrations, where $T_N$ is relatively
small, and lead to low-temperature spin-glass behavior. The
associated spin freezing would be expected to lead to a momentum
broadening of the magnetic peaks. Moreover, the observed
transition temperature would depend on the probe frequency. We
have tested for the possibility of low-temperature spin-glass
physics, and found in all samples that showed magnetic order (up
to $z=39\%$), that the width of the observed magnetic peaks
remained resolution limited. Furthermore, we measured the
temperature dependence of the magnetic scattering in a 35\%
diluted sample using incident neutron energies of 3.5, 5, and 13 meV,
with FWHM energy resolutions ranging from 0.08 to 0.72 meV,
as shown in Fig. 3. Since the transition temperature is
independent of the neutron energy resolution over a relatively wide range of
energies, we conclude that the observed behavior is a genuine
\neel~transition \cite{murani78}.

At high temperatures, pure \lco~is in the so-called
high-temperature tetragonal (HTT) phase ({\it I}4/{\it mmm}).
Below about 530 K, the oxygen octahedra surrounding the copper
ions tilt in a staggered fashion, creating an orthorhombic
distortion of the CuO$_2$ planes and forming the low-temperature
orthorhombic (LTO) phase ({\it Bmab}).  In a 8\% Zn-doped sample,
we find that the HTT to LTO transition temperature increases to
577 K, but we have not followed this transition to higher
concentrations, although previous experiments have found that it
increases monotonically with both Zn and Mg doping
\cite{cheong91}. Above $z=10\%$, we discovered a second structural
transition into a low-temperature tetragonal (LTT) phase ({\it
P}4$_2$/{\it ncm}). Figure 4 shows longitudinal scans
through the (2,0,0) position measured by neutron diffraction
above, at, and below the LTT transition temperature $T_{LTT} =
63(5) K$ for $z=19\%$.  The two peaks observed at higher temperature correspond
to the (2,0,0) and (0,2,0) reflections. Both are observed in the
same scan because of the presence of twin domains with different
orthorhombic distortion directions. In the LTT phase, only one
reflection is observed. The redistribution of intensity from the
orthorhombic peaks to the tetragonal peak without a shift in the
peak positions indicates that this is a first-order transition.
We note, that the LTT phase has previously been observed in
\lbco~\cite{axe89} and \lsnco~\cite{crawford91}.

The magnetic and structural phase diagram obtained from SQUID
magnetometry and neutron diffraction is shown in Fig. 5.  Below
$z\sim 20\%$, our data agree well with several previous results
\cite{keimer92,corti95,uchinokura95,hucker99}. Above this concentration,
we find that $T_N(z)$ deviates
from a linear behavior, approaching zero only at the percolation
threshold. Although random dilution weakens the tendency to order
\cite{vajk02}, it appears that quantum fluctuations are
not strong enough to shift the critical point: $z_{S=1/2}=z_p$,
within the uncertainty of our experiment.

The pure two-dimensional Heisenberg antiferromagnet described by
Eq. 1 can not exhibit long-range order at nonzero temperature.
However, weak inter-plane couplings and anisotropies in
\lczmo~lead to \neel~order at nonzero temperature approximately
when $\alpha_{eff}~\xi^{2}_{2D} \approx 1$, where $\xi_{2D}$
is the two-dimensional magnetic
correlation length corresponding to Eq. \ref{pure_hamiltonian}
and $\alpha_{eff}$ is a suitable combination of
the correction terms in the full spin Hamiltonian
\cite{birgeneau99}. In pure \lco, this occurs when
$\xi_{2D} \approx 100a$, where $a$ is the planar lattice constant
\cite{birgeneau99}. Using QMC results for Eq. 2 (discussed in more
detail below), we find remarkable agreement between the
$\xi_{2D}/a = 100$ contour and $T_N(z)$ up to at least 35\%.
This is demonstrated in Fig. 5a. The continuous
line indicates direct numerical results for Eq. \ref{hamiltonian},
while the dashed line results from an extrapolation
of numerical data at higher temperatures.
The full spin Hamiltonian describing \lczmo~should depend on the details
of the crystal structure. Above $z \approx 25\%$, \neel~order
occurs in the LTT phase. Nevertheless, $T_N (z)$ evolves smoothly
with doping, and corresponds to $\xi_{2D}/a =
100$ from QMC for Eq. \ref{hamiltonian} even when $T_N < T_{LTT}$,
as demonstrated in Fig. 5. Consequently, any changes with doping
in the full spin Hamiltonian must be very subtle.

A nonzero next-NN exchange in \lczmo~could shift the percolation
threshold from the NN-only value of $z_p \approx 40.7\%$.
Since the next-NN coupling in
\lczmo~is frustrating \cite{kim01}, it could, in principle, shift
the critical point noticeably to a value below the percolation
threshold (and lead to spin-glass physics, as discussed above)
\cite{birgeneau84}. However, the value of the next-NN exchange is
relatively small (0.05-0.10$J$) \cite{kim01}. For the pure
Hamiltonian Eq. 1, an additional frustrating next-NN coupling of
$\approx 0.24J$ is needed to disorder the ground state \cite{siurakshina01}.
We note that recent spin-wave measurements at the magnetic zone boundary
suggest that the dominant further-neighbor interaction is not a
next-NN exchange but a four-spin ring exchange \cite{coldea01}, which
does not extend connectivity beyond the NN percolation threshold.

\subsection{Static Structure Factor Measurement}

We have systematically studied the static structure factor in the
paramagnetic phase of \lczmo, which allowed us to determine the
instantaneous two-dimensional (2D) spin-spin correlation length $\xi(z,T)$. Above
$T_N$, the 3D magnetic Bragg peaks become rods of 2D scattering.
The equal-time structure factor $S(q_{2D})$, where $q_{2D}$ is the
2D momentum transfer component in the CuO$_2$ sheets relative to
the 2D magnetic zone center, was measured with neutron scattering
in two-axis, energy-integrating scans across these rods
\cite{birgeneau99}. Figure 6 shows representative data for two
different samples.  The measured peaks broaden as $\xi$ decreases,
both with increasing temperature and increasing dilution.
Correlation lengths were obtained from fits to a single 2D
Lorentzian
\begin{equation}
\label{lorentzian}
S(q_{2D}) = \frac{S(\pi,\pi)}{1+q_{2D}^{2}\xi^{2}}
\end{equation}
convoluted with the instrumental resolution. We note that well above
$T_N$, the spin system is effectively isotropic. However, near
$T_N$ the anisotropy and 3D coupling terms in the full spin
Hamiltonian lead to crossover physics.
Our two-axis, energy-integrating experiment
simultaneously measures both the in-plane and out-of-plane
components of the static structure factor. Since our data do not
allow for a line-shape analysis, we carried out fits to a single
2D Lorentzian.

The correlation lengths are plotted versus $J/T$ in Fig. 7a, and
the static structure factor amplitude $S(\pi,\pi)$ is shown
in Fig. 7b. Temperature is scaled by $J$ = 135 meV, the
antiferromagnetic superexchange energy of the pure system
\cite{aeppli89,tokura90}. The data are cut off above $T_N$
by one standard deviation ($\approx$ 4 K), as obtained from fits of
the order parameter (Fig. 2b). From Fig. 7a it can be seen that
dilution significantly decreases the rate at which correlations
grow as the system is cooled.  At high concentrations, $\xi(z,T)$
crosses over from exponential to power-law behavior. We note that
the $z = 40(2)\%$ and $43(2)\%$ samples do not exhibit \neel~order
at 1.4K ($J/T \approx 1100$), and the spin correlations appear to
approach a constant zero-temperature value, as expected for $z >
z_p$.

\subsection{Quantum Monte Carlo}

In order to test the degree to which \lczmo~is described by the
randomly-diluted NN Heisenberg Hamiltonian we have performed
quantum Monte Carlo (QMC) simulations to calculate $\xi(z,T)$ for
Eq. \ref{hamiltonian}.  We used a loop-cluster
algorithm \cite{vajk02,evertz93,wiese94,greven98}, with lattice
sizes 10 to 20 times larger than the correlation length in order
to avoid finite-size effects. We were able to simulate very large
lattices of up to 1700$\times$1700 sites, and to reach temperatures as
low as $T = J/100$. Previous simulations were confined to very small lattices
and high temperatures \cite{miyashita92,manousakis92}. At each
temperature and concentration, between 5 and 200 random
configurations were averaged, with 10$^4$ to 10$^5$ measurements
per configuration. The QMC results for $\xi(z,T)$, shown as black
symbols in Fig. 7a, extend to higher temperatures and complement
the experimental data. We find excellent quantitative agreement
between the two up to the percolation threshold. We emphasize
that this comparison contains no adjustable parameters, since  $J$
and $a$ are known. Above the percolation threshold, the effective
concentration of approximately 46\%
is slightly higher than the actual experimental value of 43(2)\%.
A possible origin for this might be a stronger relative influence of
the next-NN term in the full spin Hamiltonian for $z > z_p$.

In Fig. 7b, the static structure factor amplitude $S(\pi,\pi)$ from QMC
(filled symbols) is shown together with the experimental results.
Unlike the correlation length, the measured amplitude also depends
on experimental conditions, such as the effective illuminated
sample volume and the neutron flux. Since the
absolute value of $S(\pi,\pi)$ could not be determined with good
accuracy, we have normalized the experimental data for each sample
to match the respective numerical values. Note that the
temperature scale is not adjustable. The temperature dependence of
$S(\pi,\pi)$ can therefore still be compared between experiment
and QMC, and we find good agreement.

\subsection{Theory}

The ground state of the pure NN SLHAF
is ordered, but quantum fluctuations renormalize
the spin-wave velocity, $c = 2 \sqrt{2} S Z_c(S) J a$, and
spin-stiffness, $\rho_s = S^2 Z_{\rho}(S)J$, from their classical
values (using units in which $g\mu_B=k_B=\hbar=1$). For $S=1/2$,
the quantum renormalization factors $Z_c$ and $Z_{\rho}$ are known
from theoretical and numerical studies, and $\xi(z=0,T)$ is given
by \cite{chakravarty89,hasenfratz91}
\begin{equation}
\label{chn}
\frac{\xi}{a} = \frac{e}{8} \frac{c/a}{2 \pi \rho_s} e^{2 \pi \rho_s/T}
\left[1 - \frac{1}{2} \left( \frac{T}{2 \pi \rho_s} \right)
+ \mathcal{O} \left(\frac{T}{2 \pi \rho_s} \right)^2 \right]
\end{equation}
with $c=1.657Ja$ and $\rho_s = 0.18J$ \cite{beard98}. Even though
Eq. \ref{chn} is strictly valid only at asymptotically low
temperatures \cite{beard98,chakravarty89,hasenfratz91,hasenfratz00}, it agrees
remarkably well with experiment \cite{greven95,birgeneau99,kim01}
and numerics \cite{kim98,beard98} in the range $2 < \xi/a < 200$ shown
in Fig. 7a.

The derivation of Eq. \ref{chn} involves mapping the discrete
Hamiltonian (Eq. \ref{pure_hamiltonian}) to the continuum quantum
nonlinear sigma model (\qnlsm), and it assumes the existence of
an ordered ground state and translational invariance.  Random
dilution breaks translational invariance and leads to defect rods
in the Euclidian-time direction of the effective \qnlsm. These
rods break Lorentz invariance, and a \qnlsm~description may no
longer be valid \cite{chakravarty89,chernyshev02}.  Furthermore,
we can see from Fig. 7a that at higher concentrations the
temperature dependence of $\xi(z,T)$ approaches power-law
behavior. The modified form of Eq. \ref{chn}
\begin{equation}
\label{crossover}
\frac{\xi}{a} = \frac{e}{8} \frac{c/a}{2 \pi \rho_s}
\frac{e^{2 \pi \rho_s/T}}
{1 + (4 \pi \rho_s/T)^{-\nu_T}}
\end{equation}
with $\nu_T=1$ has been suggested for disorder-free systems
approaching a quantum critical point \cite{castroneto96}.  This
crossover formula interpolates between Eq. \ref{chn} at $T <
\rho_s$ and $\xi \sim 1/T$ as $\rho_s$ approaches zero. Although
recent studies indicate that $z=z_p$ is not a quantum critical
point \cite{sandvik02,sandvik02b,vajk02b}, random dilution
reduces the spin stiffness and may be viewed as bringing the
system closer to such a point in an extended parameter space.

In a classical picture, at $T=0$, $z=z_p$ is a (geometric and thermal)
multicritical point, and $\rho_s = 0$ as well as power-law
behavior of $\xi(z_p,T)$ are expected. For the $S=5/2$ system
Rb$_2$(Mn,Mg)F$_4$ at and above the percolation threshold, the
correlation length is well described by the form
\begin{equation}
\label{qd_formula}
\frac{1}{\xi(z,T)} = \frac{1}{\xi_0(z)} + \frac{1}{\xi_T(T)}
\end{equation}
where $\xi_0(z)$ is the saturated zero-temperature length and
$\xi_T(T) \sim T^{-\nu_T}$ \cite{birgeneau76}.  At the percolation
threshold, the thermal exponent of this $S=5/2$ system was found
to be $\nu_T=0.90(5)$.  A model based on the growth of 1D spin
correlations along self-avoiding walks on the percolation cluster
has been used successfully to describe these results for
Rb$_2$(Mn,Mg)F$_4$ \cite{birgeneau76} without adjustable parameters.
However, this model assumes classical spin chains, and the fact
that we obtain a different exponent for $S=1/2$ (see below) is
therefore not surprising.

We have tested the extent to which Eqs. \ref{chn} and \ref{crossover}
describe our results by fitting the numerical data using $c(z)$ and $\rho_s(z)$
as adjustable parameters. We find that fits to Eq. \ref{chn} give
a good description of $\xi(z,T)$, especially at low
concentrations, and even at $z =31\%$ for $\xi/a > 8$. However,
the modified crossover form Eq. \ref{crossover} even captures the
high-temperature power-law behavior at higher concentrations, as
shown in Fig. 7a. Results for
$c(z)$ and $\rho_s(z)$ from fits of Eq. \ref{crossover}, including
dilution levels not shown in Fig. 7, are shown in Fig. 8. Using
Eq. \ref{chn} results in large uncertainties for $c(z)$, but both
$c(z)$ and $\rho_s(z)$ obtained using the two different forms
agree within the errors. For the pure system, fits of QMC results
\cite{beard98} below $\xi/a =200$ yield $2\pi\rho_s(0) = 1.18(1)J$
and $c(0) = 1.33(3)Ja$, about 4\% higher and 20\% lower,
respectively, than the most accurate estimate \cite{beard98}.
For $z > z_p$, the combined QMC and experimental data were
fit to Eq. \ref{qd_formula}.  As can be seen in Fig. 7a, our
results are described well by this form, and we obtain
$\nu_T = 0.72(7)$ for $z=41\%$.

An attempt to combine percolation theory with the \qnlsm~model
predicts for the spin stiffness \cite{chen00}
\begin{equation}
\label{rho}
\frac{\rho_s(z)}{\rho_s(0)} = A(z) \left[1 -
\frac{\bar{g}(0)}{P_{\infty}(z)} \right]
\frac{1}{1-{\bar{g}(0)}},
\end{equation}
where $\bar{g}(0) = 0.685$ is the coupling constant corresponding
to the NN S=1/2 SLHAF at $z=0$ \cite{chakravarty89}, $A(z)$ is the
bond dilution factor, and $P_{\infty}(z)$ is the probability of a
site belonging to the infinite cluster of spins. $A(z)$ is well
described up to $z \approx 37\%$ by $A(z) \approx 1 - \pi z + \pi
z^2/2$ \cite{harris77}.  Up to $z \approx 20\%$, $P_{\infty}(z)
\approx 1-z$ \cite{sandvik02} since at low concentrations there
are very few separated clusters of spins so that $P_{\infty}(z)$
is mostly reduced from unity due to individual removed sites.
Equation \ref{rho}, shown as a dotted line in Fig. 8,
incorrectly predicts $\rho_s(z \approx 30\%) = 0$, and hence a
quantum critical point well below the percolation threshold \cite{chen00}.
To our surprise, we find that substituting $1+z$ for $1/P_{\infty}(z)$
quantitatively describes $\rho_s(z)/\rho_s(0)$ even at $z=35\%$.
This modified form of Eq. \ref{rho} is shown as a solid line in
Fig. 8. This substitution matches the original expression at low
concentrations and prevents the second term in Eq. \ref{rho} from
going to zero below the percolation threshold.  We note that Eq.
\ref{rho} is a one-loop renormalization-group result, and
(unknown) higher-order terms might perhaps improve agreement with our
observations. Recent numerical finite-size scaling results for the
spin stiffness give qualitatively similar results \cite{sandvik02},
but deviate from the unmodified form of Eq.
\ref{rho} more quickly at lower concentrations. The expression for
the spin-wave velocity corresponding to Eq. \ref{rho} is
$c(z)/c(0) = A(z)(1+z/2)$ \cite{chen00}, and is shown by the dashed line in
Fig. 8. This expression does not match our QMC fit results, possibly
for a number of reasons. Even at $z=0$, a fit results in a 20\%
error in $c$ \cite{beard98}.  Moreover, in the presence of random dilution, spin
waves become strongly damped and the spin-wave velocity is not
expected to be a well-defined quantity.

\subsection{Discussion}

Recent theoretical work for the randomly diluted $S=1/2$ NN
square-lattice Heisenberg antiferromagnet has produced exact
results at low concentration \cite{sachdev99} and led to
predictions at intermediate concentrations \cite{chen00,chernyshev02}.
Impurities may localize spin excitations and lead to a breakdown
of the classical hydrodynamic description of excitations in terms
of spin waves above a characteristic length scale
\cite{chernyshev02}. Static properties such as the staggered
magnetization and $T_N$, as well as the spin stiffness $\rho_s$,
are expected to remain well defined.
On the other hand, dynamic observables, such as the spin-wave
velocity $c$, become ill-defined in this picture. As long as the
ground state remains ordered, the low-temperature correlation
length should still scale as $\xi \sim e^{2\pi\rho_s/T}$.  This is
consistent with our observation for $\xi(z,T)$ over a wide range
of concentrations and temperatures, which we find to be well
described by Eq. \ref{chn}, and especially the heuristic crossover
form Eq. \ref{crossover}.

Although the percolation transition for the 2D S=1/2 Heisenberg
antiferromagnet may be classical \cite{sandvik02,sandvik02b,vajk02b},
recent QMC simulations of diluted bilayers indicate that the percolation
threshold for the single layer system is very close to a new
quantum multi-critical point in an extended parameter space
\cite{sandvik02b,vajk02b}. Indeed, the dynamic critical
exponent obtained in these studies is consistent with the
thermal correlation length exponent $\nu_T \approx 0.7$ that we have
found here. Therefore, in the temperature regime that we have studied,
the properties of \lczmo~near the percolation threshold appear to be
controlled by this new critical point.
Future inelastic neutron scattering
measurements of the full dynamic structure factor should give
further insight into this complex quantum many-body system.

O.P.V. and M.G. thank A. Aharony, A. H. Castro Neto, A. L.
Chernyshev, A. W. Sandvik, and E. F. Shender for helpful
discussions. The work at Stanford was supported by the U.S.
Department of Energy under contract nos. DE-FG03-99ER45773 and
DE-AC03-76SF00515, by NSF CAREER Award no. DMR9985067, and
by the A.P. Sloan Foundation.


\begin{figure}
\label{spins}
\includegraphics[width=8.5cm]{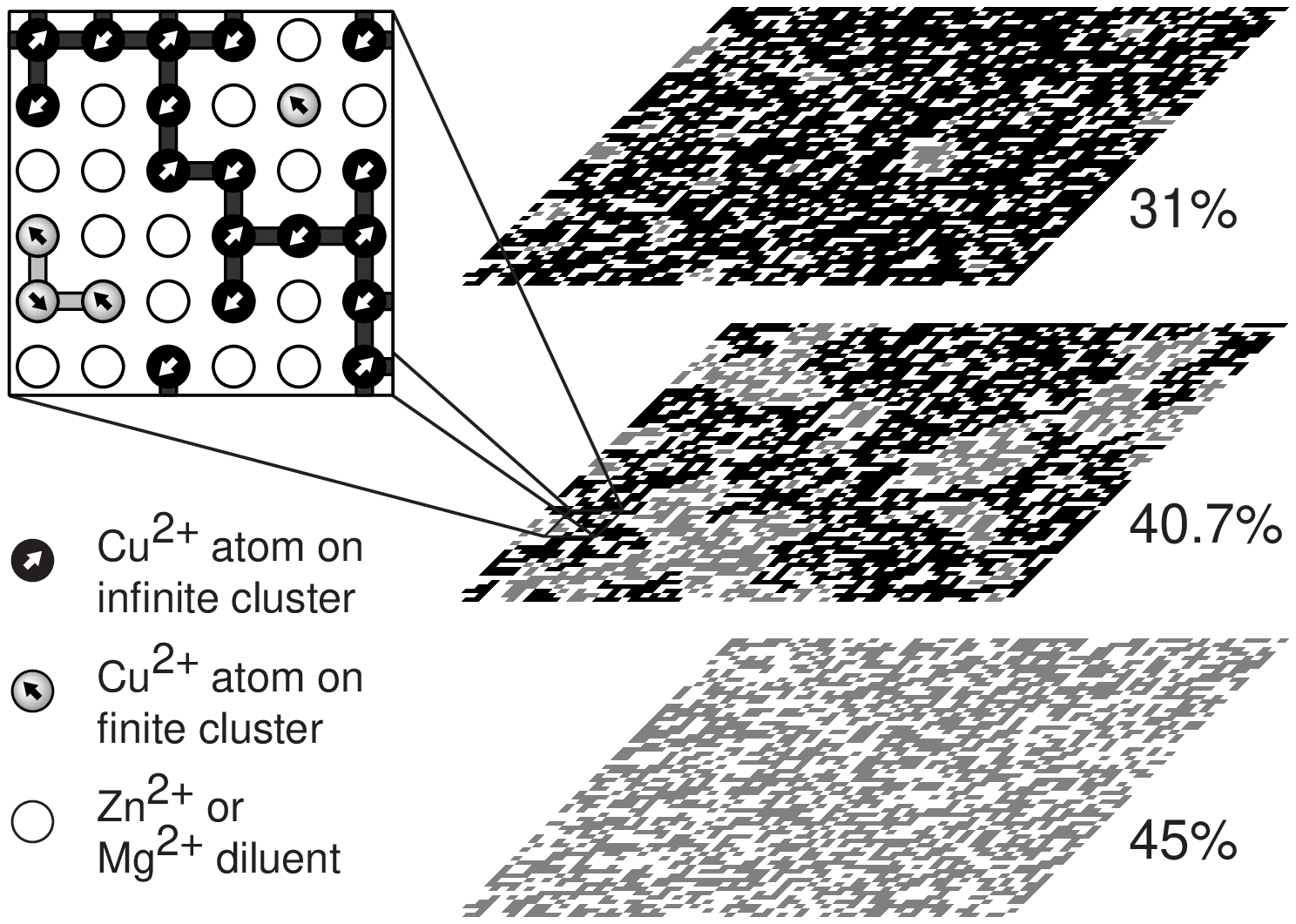}
\caption{
        Schematic of finite-sized sections of the infinite square lattice with random
        site dilution.  Sites on the infinite cluster are shown in black, sites on finite
        disconnected clusters in grey, and diluents in white.  At a concentration of
        31\%, most of the lattice is still connected.  At 40.7\%,
        just below the percolation threshold, much of the lattice is disconnected,
        but there is still a percolating cluster that spans the infinite
        lattice.  Above the percolation threshold, all clusters are of finite size.  The
        inset is a close-up view of the lattice, showing the role that magnetic Cu and non-magnetic
        Zn/Mg ions play in the experimental system.
        }
\end{figure}

\begin{figure}
\label{neel}
\includegraphics[width=8.5cm]{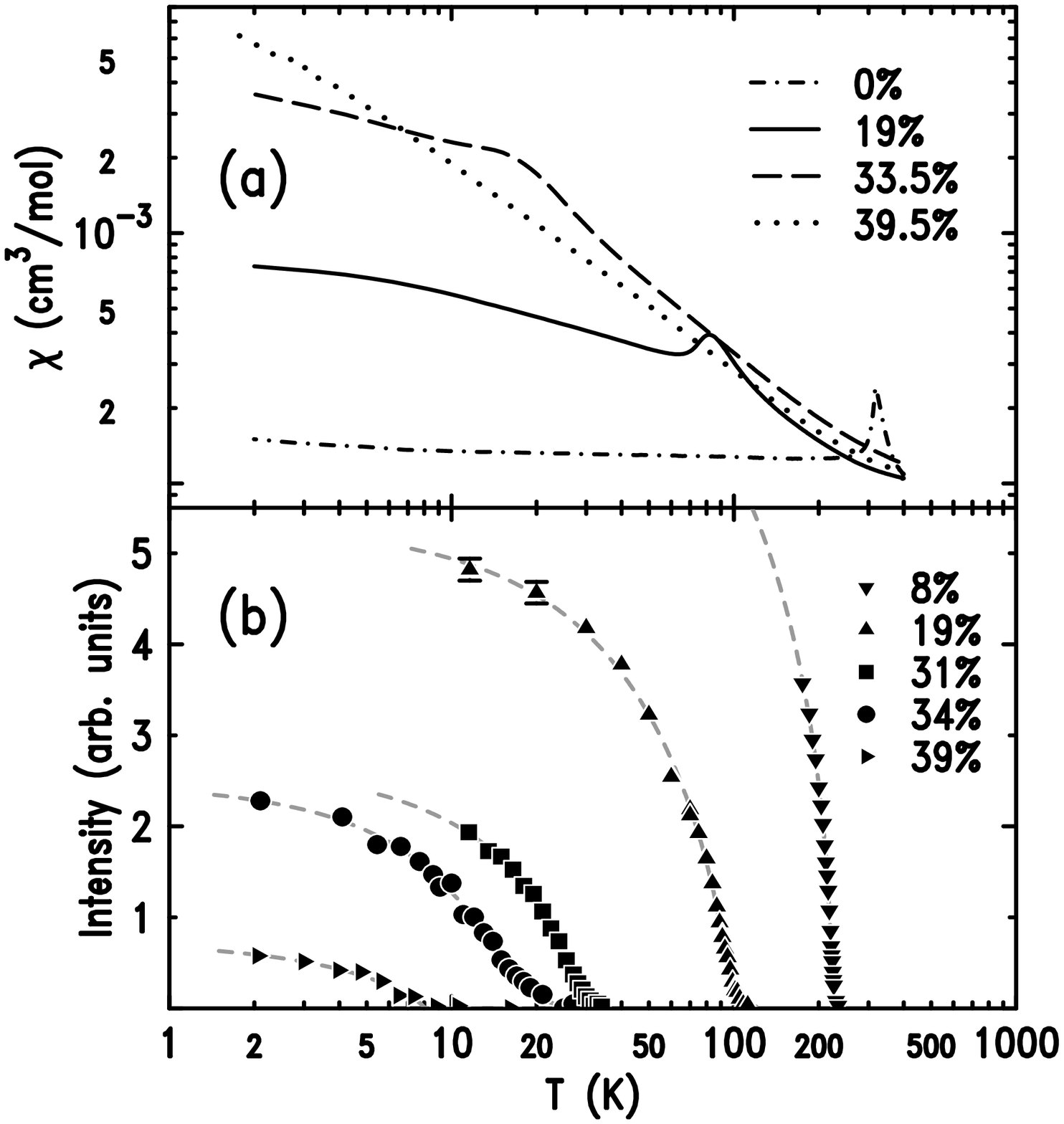}
\caption{
        Measurements of the \neel~temperature $T_N (z)$ for various samples.
        a) Magnetic susceptibility measured with a SQUID magnetometer
        with a 500 Oe field applied along the $c$ axis.  The
        susceptibility exhibits a cusp at $T_N$.  At higher concentrations,
        a strong Curie-like component emerges, making the cusps more
        difficult to distinguish.
        b) (1,0,0) magnetic Bragg peak intensity from neutron diffraction.
        A temperature-independent component, which is predominantly due to
        nuclear double scattering, has been subtracted.
        The lines represent fits for the magnetic order parameter
        squared, $\sim (T_N - T)^{2\beta}$ with $\beta \approx 0.30$,
        assuming a Gaussian distribution
        of $T_N$ (typically $\approx$ 4 K) due to inhomogeneities present
        in the large samples used.
        }
\end{figure}

\begin{figure}
\label{spinglass}
\includegraphics[width=8.5cm]{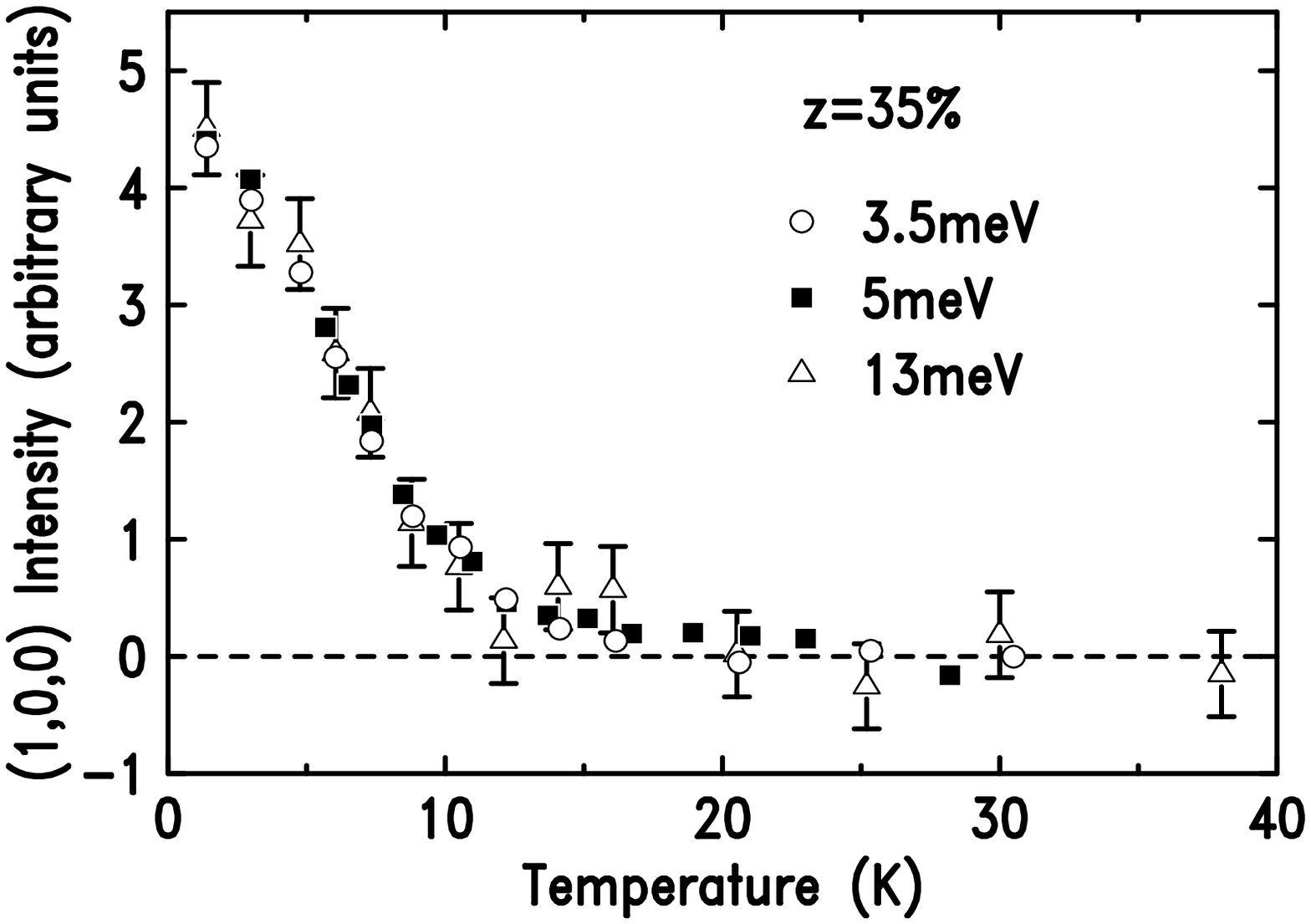}
\caption{
        Order parameter measurements performed on the 35\%
        sample on the cold neutron instrument using neutron energies of 3.5, 5, and 13 meV.
        The corresponding energy resolutions range from 0.08 to 0.72 meV (FWHM).
        Even at very high concentrations close to the percolation threshold,
        the order parameter measurement remains independent of the energy resolution,
        indicating \neel~rather than spin-glass order.  A temperature-independent
        component of the signal, which is highly energy-dependent and predominantly
        due to double scattering involving the nuclear (1,0,2)
        and (0,0,2) peaks, has been subtracted from the data.
        }
\end{figure}

\begin{figure}
\label{LTT}
\includegraphics[width=8.5cm]{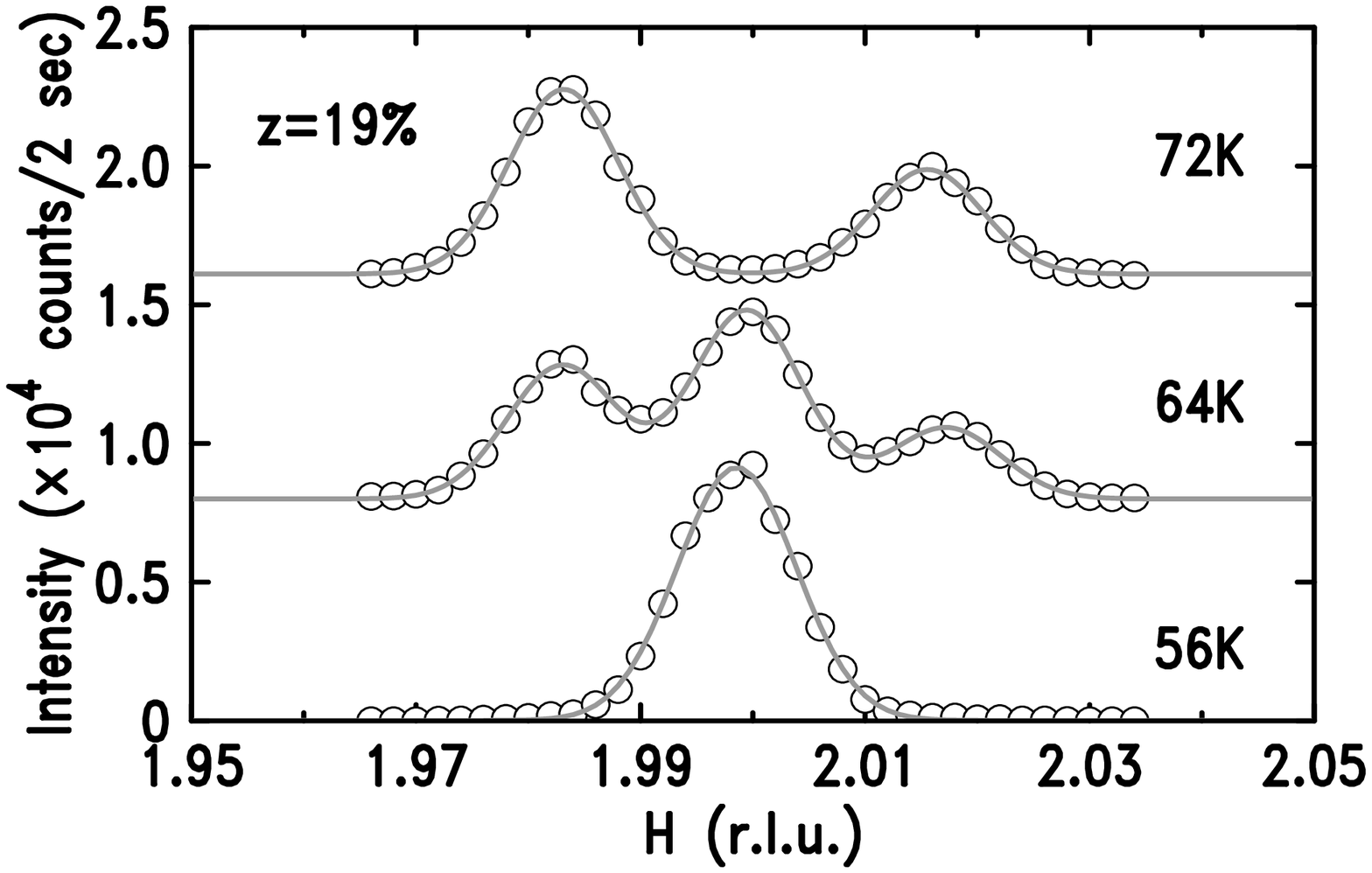}
\caption{
        Neutron diffraction measurements of the first-order
        structural phase transition from the orthorhombic
        to low-temperature tetragonal phase at 19\%
        dilution using 14.7 meV neutrons and collimations of 10'-27.5'-sample-23.7'-open.
        Scans are offset vertically, and lines indicate Gaussian fits.
        Above the transition, the longitudinal scans show a superposition of
        (2,0,0) and (0,2,0) peaks from different crystal domains.
        In the transition region, both phases coexist, partly because of the
        weakly-first-order nature of the transition, and because of the small
        concentration gradient present in our samples.
        In the tetragonal phase, only one peak is observed.
        }
\end{figure}

\begin{figure}
\label{phase}
\includegraphics[width=8.5cm]{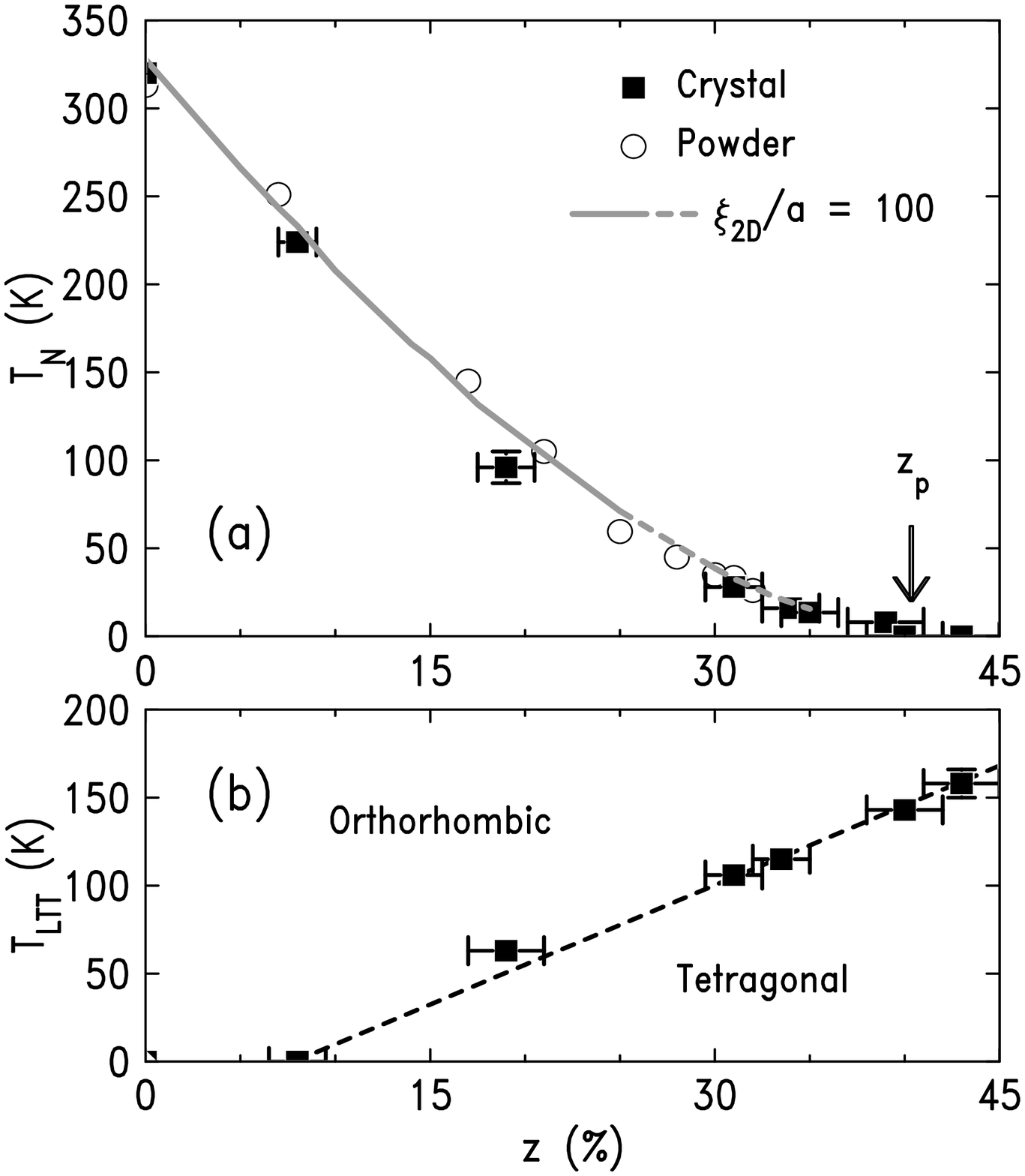}
\caption{
        Magnetic and structural phase diagram of \lczmo.
        a) \neel~temperature vs. dilution.
        Single crystal results are from neutron measurements
        of the order parameter, powder results are from SQUID
        susceptibility measurements of polycrystalline samples.
        Our results up to $z = 25\%$ are in good agreement with
        previous work.
        The decrease of $T_N (z)$ at higher concentrations is not linear, but
        more gradual. Within the uncertainty of our experiment,
        $T_N (z)$ reaches zero at the percolation threshold,
        $z_p \approx 40.7\%$.
        b) Structural transition temperature from orthorhombic to
        low-temperature tetragonal phase, as measured by neutron
        diffraction. The doping dependence of the transition
        temperature is approximately linear, as indicated by the
        dashed line.
        }
\end{figure}

\begin{figure}
\label{2axis}
\includegraphics[width=8.5cm]{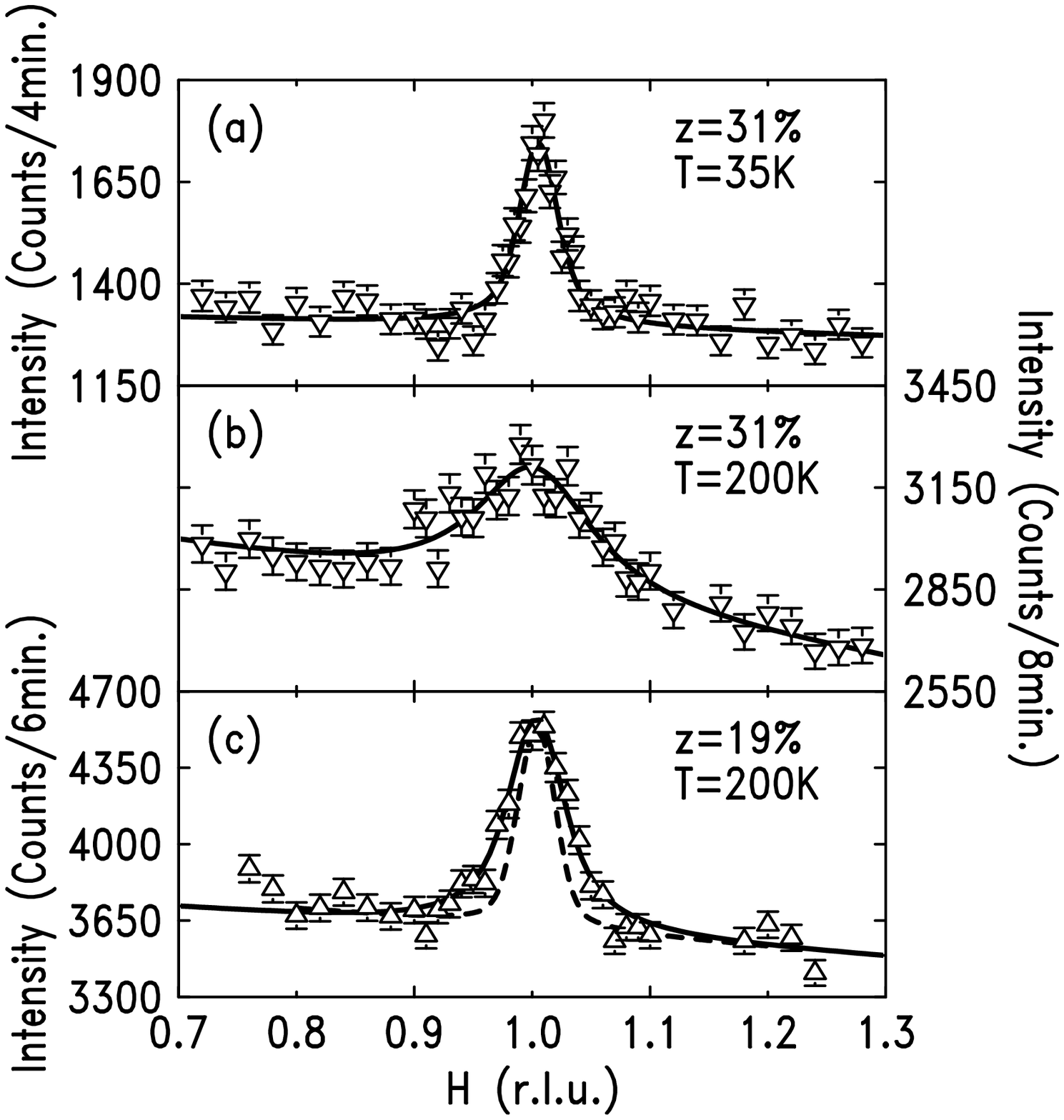}
\caption{
        Two-axis, energy-integrating measurements of the static
        structure factor taken with 30.5 meV incident neutron
        energy and horizontal collimations of
        40'-27.5'-sample-23.7'.
        At 31\% dilution,
        a) just above $T_N$ and
        b) well above $T_N$.
        c) At 19\% dilution, just above $T_N$.
        The dashed line in c) indicates the instrumental resolution.
        Solid lines show fits to a 2D Lorentzian convoluted with the resolution, as
        discussed in the text.
        }
\end{figure}

\begin{figure}
\label{xi_amp}
\includegraphics[width=8.5cm]{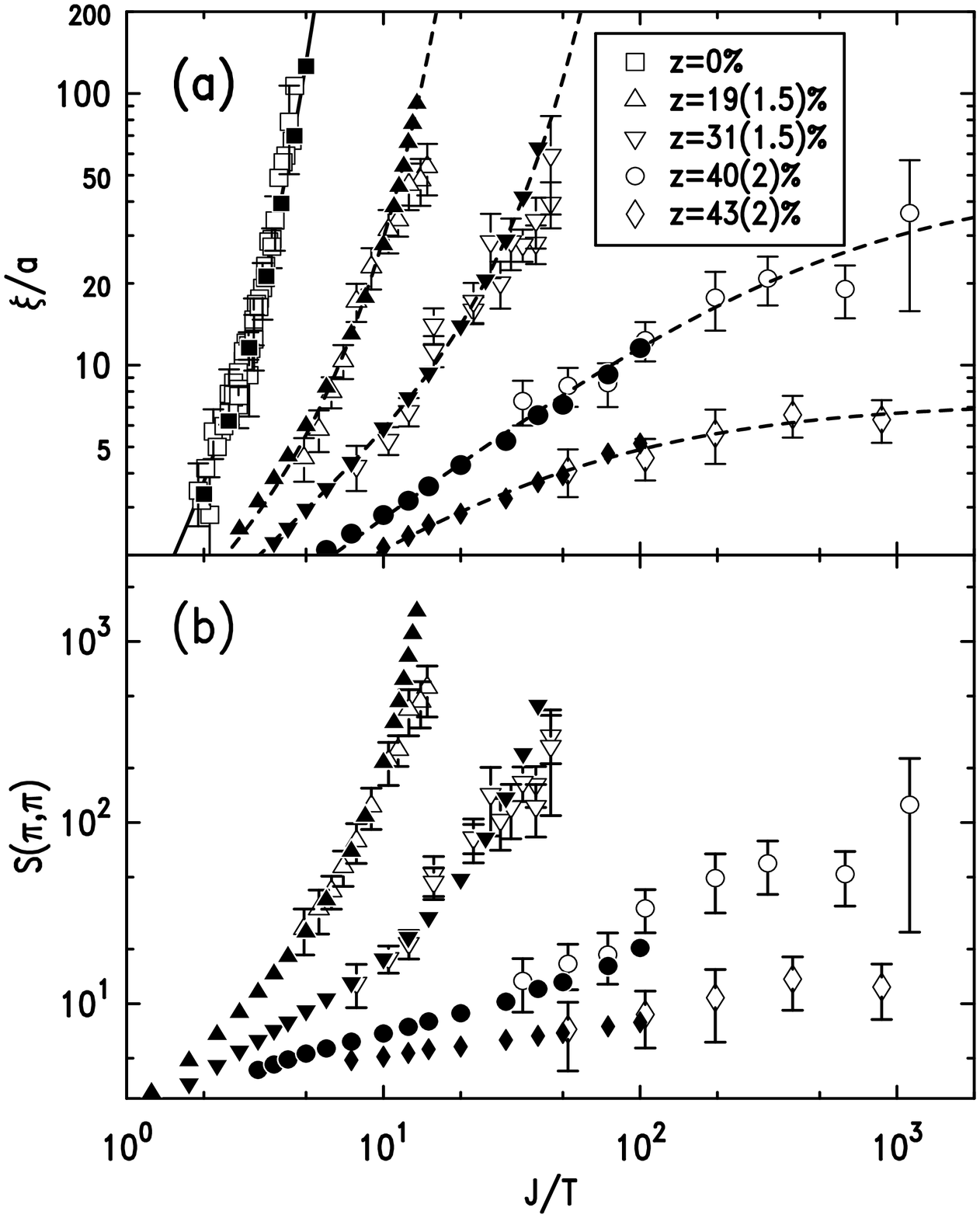}
\caption{
        a)  Spin-spin correlation lengths, in units of the lattice constant,
        versus inverse temperature, in units of the NN superexchange
        $J = 135$ meV of the pure system.  Open symbols represent results from
        neutron scattering measurements of \lczmo; filled symbols represent
        quantum Monte Carlo (QMC) data for z=0, 20, 31, 41, and 46\%.
        No adjustable parameters were used in the comparison.  Experimental and QMC
        results for $z=0$ are from previous work \cite{birgeneau99,beard98}.
        The dashed lines are fits to Eqs. \ref{crossover} and \ref{qd_formula},
        as described in the text, and the solid line is Eq. \ref{chn}.
        b)  Static structure factor peak amplitude versus inverse temperature.
        Open symbols are neutron scattering results and filled symbols are from QMC,
        as above.  Normalization between experimental and numerical results
        is discussed in the text.
        }
\end{figure}

\begin{figure}
\label{fit_results}
\includegraphics[width=8.5cm]{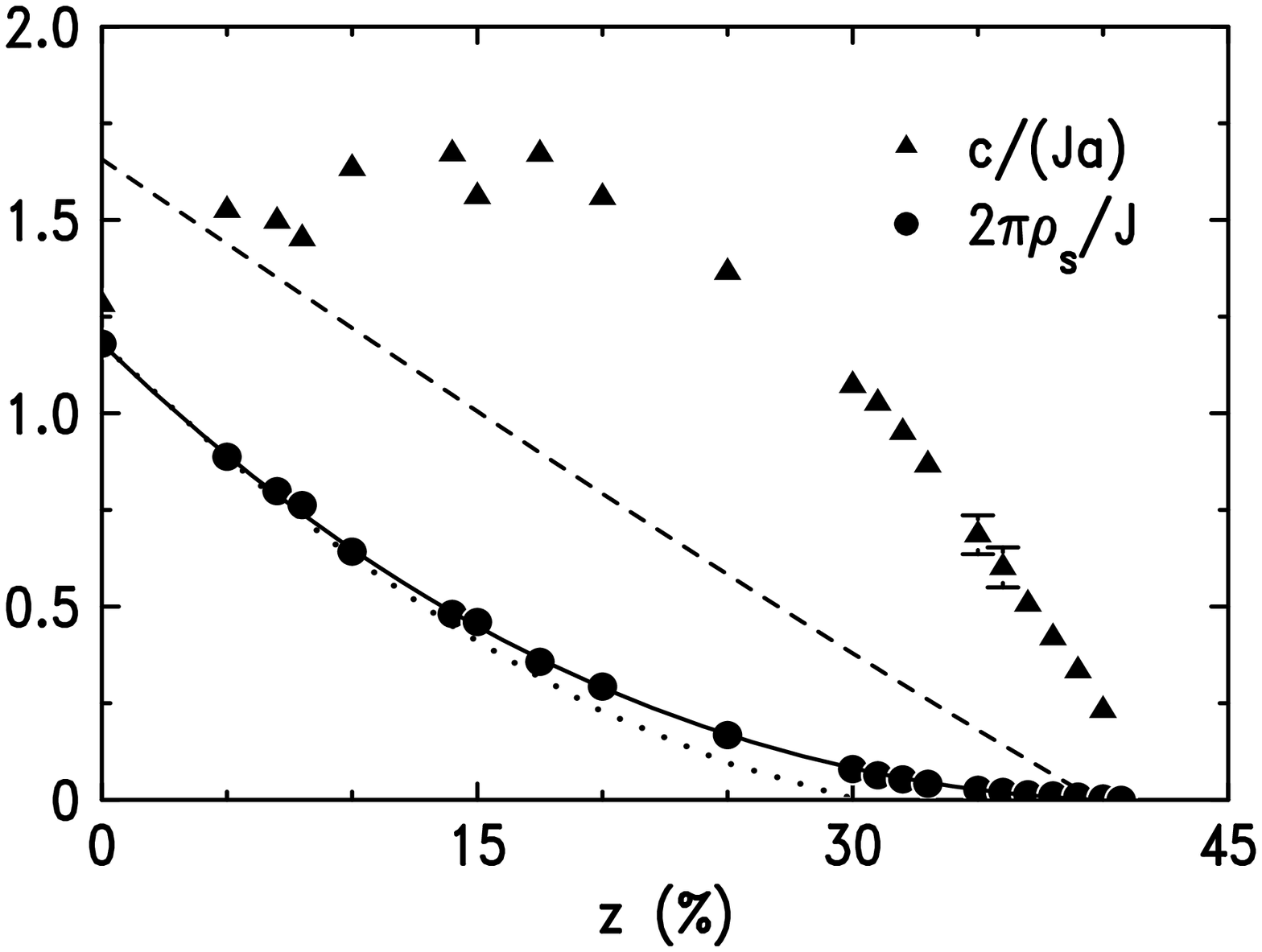}
\caption{
        Effective spin-wave velocity $c(z)$ and spin stiffness $2\pi\rho_s (z)$
        as a function of non-magnetic concentration $z$
        extracted from fits to Monte Carlo results using
        the heuristic crossover form Eq. \ref{crossover}.  Lines are
        discussed in the text.
        }
\end{figure}


\bibliography{greven}

\begin{thebibliography}{10}
\expandafter\ifx\csname url\endcsname\relax
  \def\url#1{\texttt{#1}}\fi
\expandafter\ifx\csname urlprefix\endcsname\relax\def\urlprefix{URL }\fi

\bibitem{lyons88}
K.~B. Lyons, P.~A. Fleury, J.~P. Remeika, A.~S. Cooper, T.~J. Negran, Phys.
  Rev. B 37 (1988) 2353.

\bibitem{endoh88}
Y.~Endoh, K.~Yamada, R.~J. Birgeneau, D.~R. Gabbe, H.~P. Jennsen, M.~Kastner,
  Phys. Rev. B 37 (1988) 7443.

\bibitem{aeppli89}
G.~Aeppli, S.~M. Hayden, H.~A. Mook, Z.~Fisk, S.-W. Cheong, D.~Rytz, J.~P.
  Remeika, G.~P. Espinosa, A.~S. Cooper, Phys. Rev. Lett. 62 (1989) 2052.

\bibitem{tokura90}
Y.~Tokura, S.~Koshihara, T.~Arima, H.~Takagi, S.~Ishibashi, T.~Ido, S.~Uchida,
  Phys. Rev. B 41 (1990) 11657.

\bibitem{keimer92}
B.~Keimer, N.~Belk, R.~J. Birgeneau, A.~Cassanho, C.~Y. Chen, M.~Greven, M.~A.
  Kastner, A.~Aharony, Y.~Endoh, R.~W. Erwin, G.~Shirane, Phys. Rev. B 46
  (1992) 14034.

\bibitem{greven95}
M.~Greven, R.~J. Birgeneau, Y.~Endoh, M.~A. Kastner, M.~Matsuda, G.~Shirane, Z.
  Phys. B 96 (1995) 465.

\bibitem{birgeneau99}
R.~J. Birgeneau, M.~Greven, M.~A. Kastner, Y.~S. Lee, B.~O. Wells, Phys. Rev. B
  59 (1999) 13788.

\bibitem{kim01}
Y.~J. Kim, R.~J. Birgeneau, F.~C. Chou, M.~Greven, M.~A. Kastner, Y.~S. Lee,
  B.~O. Wells, A.~Aharony, O.~Entin-Wohlman, I.~Y. Korenblit, A.~B. Harris,
  Phys. Rev. B 64 (2001) 024435.

\bibitem{kim01L}
Y.~J. Kim, R.~J. Birgeneau, F.~C. Chou, R.~W. Erwin, M.~A. Kastner, Phys. Rev.
  Lett. 86 (2001) 3144.

\bibitem{coldea01}
R.~Coldea, S.~M. Hayden, G.~Aeppli, T.~G. Perring, C.~D. Frost, T.~E. Mason,
  S.~W. Cheong, Z.~Fisk, Phys. Rev. Lett. 86 (2001) 5377.

\bibitem{makivic91}
M.~Makivi\'c, H.~Q. Ding, Phys. Rev. B 43 (1991) 3562.

\bibitem{kim98}
J.-K. Kim, M.~Troyer, Phys. Rev. Lett. 80 (1998) 2705.

\bibitem{beard98}
B.~B. Beard, R.~J. Birgeneau, M.~Greven, U.-J. Wiese, Phys. Rev. Lett. 80
  (1998) 1742.

\bibitem{chakravarty89}
S.~Chakravarty, B.~I. Halperin, D.~R. Nelson, Phys. Rev. B 39 (1989) 2344.

\bibitem{hasenfratz91}
P.~Hasenfratz, F.~Niedermayer, Phys. Lett. B 268 (1991) 231.

\bibitem{tyc89}
S.~Ty\v{c}, B.~I. Halperin, S.~Chakravarty, Phys. Rev. Lett 62 (1989) 835.

\bibitem{singh89}
R.~R.~P. Singh, Phys. Rev. B 39 (1989) 9760.

\bibitem{igarashi92}
J.~Igarashi, Phys. Rev. B 46 (1992) 10763.

\bibitem{chubukov94}
A.~V. Chubukov, S.~Sachdev, J.~Ye, Phys. Rev. B 49 (1994) 11919.

\bibitem{hamer94}
C.~J. Hamer, Z.~Weihong, J.~Oitmaa, Phys. Rev. B 50 (1994) 1994.

\bibitem{elstner95}
N.~Elstner, A.~Sokol, R.~R.~P. Singh, M.~Greven, R.~J. Birgeneau, Phys. Rev.
  Lett. 75 (1995) 938.

\bibitem{singh95}
R.~R.~P. Singh, M.~P. Gelfand, Phys. Rev. B 52 (1995) 15695.

\bibitem{cuccoli97}
A.~Cuccoli, V.~Tognetti, R.~Vaia, P.~Verrucchi, Phys. Rev. B 56 (1997) 14456.

\bibitem{hasenfratz00}
P.~Hasenfratz, Eur. Phys. J. B 13 (2000) 11.

\bibitem{siurakshina01}
L.~Siurakshina, D.~Ihle, R.~Hayn, Phys. Rev. B 64 (2001) 104406.

\bibitem{shevchenko00}
P.~V. Shevchenko, A.~W. Sandvik, O.~P. Sushkov, Phys. Rev. B 61 (2000) 3475.

\bibitem{annett89}
J.~F. Annett, R.~M. Martin, A.~K. McMahan, S.~Satpathy, Phys. Rev. B 40 (1989)
  2620.

\bibitem{young96}
A.~P. Young, H.~Rieger, Phys. Rev. B 53 (1996) 8486.

\bibitem{sachdev97}
S.~Sachdev, A.~P. Young, Phys. Rev. Lett. 78 (1997) 2220.

\bibitem{senthil96}
T.~Senthil, S.~Sachdev, Phys. Rev. Lett. 77 (1996) 5292.

\bibitem{heuer92}
H.-O. Heuer, Phys. Rev. B 45 (1992) 5691.

\bibitem{sandvik97}
A.~W. Sandvik, E.~Dagotto, D.~J. Scalapino, Phys. Rev. B 56 (1997) 11701.

\bibitem{sachdev99}
S.~Sachdev, C.~Buragohain, M.~Vojta, Science 286 (1999) 2479.

\bibitem{vojta00}
M.~Vojta, C.~Buragohain, S.~Sachdev, Phys. Rev. B 61 (2000) 15152.

\bibitem{chernyshev02}
A.~L. Chernyshev, Y.~C. Chen, A.~H. {Castro Neto}, Phys. Rev. B 65 (2002)
  104407.

\bibitem{yasuda99}
C.~Yasuda, A.~Oguchi, J. Phys. Soc. Japan 68 (1999) 2773.

\bibitem{chen00}
Y.-C. Chen, A.~H. {Castro Neto}, Phys. Rev. B 61 (2000) R3772.

\bibitem{sandvik95}
A.~W. Sandvik, M.~Veki\'c, Phys. Rev. Lett. 74 (1995) 1226.

\bibitem{birgeneau84}
R.~J. Birgeneau, R.~A. Cowley, G.~Shirane, H.~Yoshizawa, J. Stat. Phys. 34
  (1984) 817, and references therein.

\bibitem{chakraborty89}
A.~Chakraborty, A.~J. Epstein, M.~Jarrel, E.~M. McCarron, Phys. Rev. B 40
  (1989) 5296.

\bibitem{cheong91}
S.-W. Cheong, A.~S. Cooper, L.~W.~J. Rupp, B.~Batlogg, Phys. Rev. B 44 (1991)
  9739.

\bibitem{ting92}
S.~T. Ting, P.~Pernambuco-Wise, J.~E. Crow, E.~Manousakis, Phys. Rev. B 46
  (1992) 11772.

\bibitem{lichti91}
R.~L. Lichti, C.~Boekema, J.~C. Lam, D.~W. Cooke, S.~F.~J. Cox, S.~T. Ting,
  J.~E. Crow, Physica C 180 (1991) 358.

\bibitem{cao94}
G.~Cao, J.~W. O'Reilly, J.~E. Crow, L.~R. Testardi, J. Appl. Phys. 75 (1994)
  6595.

\bibitem{corti95}
M.~Corti, A.~Rigamonti, F.~Tabak, P.~Carretta, F.~Licci, L.~Raffo, Phys. Rev. B
  52 (1995) 4226.

\bibitem{uchinokura95}
K.~Uchinokura, T.~Ino, I.~Terasaki, I.~Tsukada, Physica B 205 (1995) 234.

\bibitem{hucker99}
M.~H\"{u}cker, V.~Kataev, J.~Pommer, J.~Harras, A.~Hosni, C.~Pflitsch,
  R.~Gross, B.~B\"{u}chner, Phys. Rev. B 59 (1999) R725.

\bibitem{clarke95}
S.~J. Clarke, A.~Harrison, J. Mag. Mag. Materials 140 (1995) 1627.

\bibitem{breed70}
D.~J. Breed, K.~Gilijamse, J.~W.~E. Sterkenburg, A.~R. Miedema, J. Appl. Phys.
  41 (1970) 1267.

\bibitem{Stauffer}
D.~Stauffer, A.~Aharony, Introduction to Percolation Theory, Revised $2^{\rm
  nd}$ Ed., Taylor \& Francis, Bristol, PA, 1994.

\bibitem{newman00}
M.~E.~J. Newman, R.~M. Ziff, Phys. Rev. Lett. 85 (2000) 4104.

\bibitem{miyashita92}
S.~Miyashita, J.~Behre, S.~Yamamoto, in: Computational Approaches in Condensed
  Matter Physics, Springer, Berlin, 1992, p.~97.

\bibitem{kato00}
K.~Kato, S.~Todo, K.~Harada, N.~Kawashima, S.~Miyashita, H.~Takayama, Phys.
  Rev. Lett. 84 (2000) 4204.

\bibitem{todo01}
S.~S. Todo, H.~Takayama, N.~Kawashima, Phys. Rev. Lett 86 (2001) 3210.

\bibitem{yasuda01}
C.~Yasuda, S.~Todo, M.~Matsumoto, H.~Takayama, Phys. Rev. B 64 (2001) 092405.

\bibitem{vajk02}
O.~P. Vajk, P.~K. Mang, M.~Greven, P.~M. Gehring, J.~W. Lynn, Science 295
  (2002) 1691.

\bibitem{sandvik02}
A.~W. Sandvik, Phys. Rev. B 66 (2002) 024418.

\bibitem{sandvik02b}
A.~W. Sandvik, Phys. Rev. Lett. 89 (2002) 177201.

\bibitem{vajk02b}
O.~P. Vajk, M.~Greven, Phys. Rev. Lett. 89 (2002) 177202.

\bibitem{aharony88}
A.~Aharony, R.~J. Birgeneau, A.~Coniglio, M.~A. Kastner, H.~E. Stanley, Phys.
  Rev. Lett. 60 (1988) 1330.

\bibitem{thio88}
T.~Thio, T.~R. Thurston, N.~W. Preyer, P.~J. Picone, M.~A. Kastner, H.~P.
  Jennsen, D.~R. Gabbe, C.~Y. Chen, R.~J. Birgeneau, A.~Aharony, Phys. Rev. B
  38 (1988) 905.

\bibitem{murani78}
A.~P. Murani, A.~Heidemann, Phys. Rev. Lett. 41 (1978) 1402.

\bibitem{axe89}
J.~D. Axe, A.~H. Moudden, D.~Hohlwein, D.~E. Cox, K.~M. Mohanty, A.~R.
  Moodenbaugh, Y.~Xu, Phys. Rev. Lett. 62 (1989) 2751.

\bibitem{crawford91}
M.~K. Crawford, R.~L. Harlow, E.~M. McCarron, W.~E. Farneth, J.~D. Axe,
  H.~Chou, Q.~Huang, Phys. Rev. B 44 (1991) 7749.

\bibitem{evertz93}
H.~G. Evertz, G.~Lana, M.~Marcu, Phys. Rev. Lett. 70 (1993) 875.

\bibitem{wiese94}
U.-J. Wiese, H.-P. Ying, Z. Phys. B 93 (1994) 147.

\bibitem{greven98}
M.~Greven, R.~J. Birgeneau, Phys. Rev. Lett. 81 (1998) 1945.

\bibitem{manousakis92}
E.~Manousakis, Phys. Rev. B 45 (1992) 7570.

\bibitem{castroneto96}
A.~H. {Castro Neto}, D.~Hone, Phys. Rev. Lett. 76 (1996) 2165.

\bibitem{birgeneau76}
R.~J. Birgeneau, R.~A. Cowley, G.~Shirane, H.~J. Guggenheim, Phys. Rev. Lett.
  37 (1976) 940.

\bibitem{harris77}
A.~B. Harris, S.~Kirkpatrick, Phys. Rev. B 16 (1977) 542.

\end{thebibliography}

\end{document}